\documentclass[prl,twocolumn,superscriptaddress]{revtex4}

\usepackage{graphicx}

\begin{document}

\title{Gossamer Superconductor, Mott Insulator, and Resonanting Valence Bond State in Correlated Electron Systems}
\author{F. C. Zhang}
\affiliation{Department of Physics, University of Cincinnati, Cincinnati OH 45221\\
and\\
Los Alamos National Laboratory, Los Alamos, NM 87545}

\begin{abstract}
Gutzwiller variational method is applied to an effective two-dimensional Hubbard model
to examine the recently proposed gossamer superconductor by Laughlin\cite{laughlin}.
The ground state at half filled electron density
is a gossamer superconductor for smaller intra-site
Coulomb repulsion $U$ and a Mott insulator for larger $U$.
The gossamer superconducting state is similar to the resonant valence bond
superconducting state, except that the chemical potential
is approximately pinned at the mid of the two Hubbard bands
away from the half filled.

\end{abstract}


\maketitle

Theories for  high temperature superconductivity  continue to attract much interest
in condensed matter physics.
Soon after its discovery, Anderson proposed the idea of
resonant valence bond (RVB) state for the observed unusual properties in high $T_c$
superconducting Cu-oxides\cite{anderson}. In the
RVB picture, each lattice site is either unoccupied or singly occupied by a spin-up or down electron.
The spins are coupled antiferromagnetically without long range order. The charge carriers move in the
spin liquid background and condense to a superconducting state\cite{anderson2,palee}.
The RVB states are oftenly studied using 2-dimensional Hubbard or
t-J models~\cite{anderson,zhang1}. In this scenario,
the undoped cuprate with density one electron per site, or half filled, is a Mott insulator,
and the superconductor is viewed as a doped Mott insulator when additional holes or electrons are introduced.
A Mott insulator is a special type of insulator casued by electron interaction.
It has been established that the ground state of many models in 1-dimensional chain or in ladders
at half filled are Mott insulator\cite{dagotto}. In two or higher dimensions,
a Mott insulator has
a strong tendency toward antiferromagnetic or other types of ordering states breaking
translational symmetry.

Very recently, Laughlin has proposed an interesting new notion, the gossamer superconductor, for high temperature
superconducting Cu-oxides\cite{laughlin}.
In a gossamer superconductor, the superfluid density is very thin, in contrast to
the conventional superconductor.  Laughlin has proposed an explicit microscopic
wavefunction for the gossamer superconductor, which has the following form,
\begin{eqnarray}
\mid \Psi_L \rangle = \prod\limits_i(1-\alpha n_{i \uparrow}n_{i \downarrow}) \mid \Psi_{BCS} \rangle \\
\mid \Psi_{BCS} \rangle =
\prod\limits_{\vec k} ( u_{\vec k} + v_{\vec k} c^{\dag}_{\vec k \uparrow} c^{\dag}_{-\vec k \downarrow})
\mid 0 \rangle \nonumber
\end{eqnarray}
where $\mid \Psi_{BCS} \rangle $ is a BCS superconducting state,
and $\Pi_{\alpha}=\prod_i (1-\alpha n_{i \uparrow}n_{i \downarrow})$
is a projection operator to partially project out doubly occupied electron states on each lattice
site $i$.  $n_{i\sigma}=c^{\dagger}_{i\sigma}c_{i\sigma}$ is the electron number operator
of spin $\sigma = \uparrow$, $\downarrow$
at site $i$, and $\alpha$ is a parameter between 0 and 1.
If $\alpha$ is close to 1, the projection operator strongly suppresses
the superfluid density.
In the gossamer superconducting state, the probability to have both spin-up and spin-down electrons occupying the
same lattice site is largely reduced but remains to be finite.  Because of the partial projection,
the state is superconducting even at  half filling.  This is
different from the RVB theory where the
electron doubly occupied states are completely projected out, hence
the half filled RVB state is a Mott insulator and the
superconductivity occurs only away from the half filled.  As it has been shown by
Laughlin\cite{laughlin},
the gossamer superconducting state in Eq. (1) is an exact ground state of a model Hamiltonian given by
$H_L = \sum_{\vec k \sigma} E_{\vec k \sigma} \tilde{b}^{\dagger}_{\vec k \sigma} \tilde{b}_{\vec k \sigma}$,
where $E_{\vec k \sigma} \geq 0$,
$\tilde{b}_{\vec k \sigma} = \Pi_{\alpha} b_{\vec k \sigma}\Pi^{-1}_{\alpha}$,
and $b_{\vec k \sigma}$ is the quasiparticle annihilation operator of the BCS state,
$b_{\vec k \sigma} \mid \Psi_{BCS} \rangle =0$. $\Pi^{-1}_{\alpha}$
is the inverse of $\Pi_{\alpha}$.
Laughlin has also
argued that the gossamer superconducting state is related  to the large on-site Coulomb repulsion.                                   \

It will be interesting to examine the possible gossamer superconducting state in a more realistic model.
As it is generally believed that the large Coulomb repulsion may lead to a Mott insulator
at half filling, it will also be interesting to examine
the possibility of the phase transition from a Mott insulator to a gossamer superconductor
as the electron interaction strength decreases. Since a doped Mott insulator
can be a RVB superconducting state, it is natural to ask the question of
the similarities and the differences between the gossamer and the  RVB superconductors.

In this paper we intend to examine these questions by studying an effective Hubbard model given in Eq. (2)
below using
Gutzwiller's variational method~\cite{gutzwiller}.  In the Gutzwiller's approach,
the on-site Coulomb repulsion is treated exactly, while the kinetic energy is studied
variationally, so that it is suitable to examine some issues in strongly correlated systems.
That method was used by Brinkman and Rice\cite{brinkman}
to study the phase transition between an insulator and a metallic state described by
a partially projected Fermi liquid state.  The variational method applied to
the effective Hubbard model in two-dimension demonstrates a phase transition
from a gossamer superconductor for smaller intra-site
Coulomb repulsion $U$ to a Mott insulator for larger $U$ at half filling.
The gossamer superconductor
is shown similar to the RVB superconducting state of the doped Mott insulator. However, its chemical
potential is found to be approximately pinned at the mid of the lower and the higher Hubbard
bands, different from the RVB state where the chemical potential is shifted to the lower Hubbard band
upon doping.

We study an effective Hubbard Hamiltonian,
\begin{eqnarray}
H  = U \sum\limits_i n_{i \uparrow} n_{i \downarrow}-\sum\limits_{\langle i j \rangle \sigma}
(t_{ij} c^{\dag}_{i \sigma} c_{j \sigma}+h.c.)
 +\sum\limits_{\langle ij \rangle} J_{ij}\vec S_i \cdot \vec S_j
\end{eqnarray}
In this Hamiltonian,  we have
introduced an
antiferromagnetic spin-spin coupling term ($J_{ij} \geq 0$)
 to account for the virtual electron double occupancy
effect. In the large $U$ limit, $J_{ij} \approx 4t_{ij}^2/U$.  This model may be viewed as an effective
Hamiltonian of the Hubbard model.  The inclusion of the antiferromagnetic spin coupling
appears consistent with the weak coupling
renormalization group analyses~\cite{honerkamp}, and is appropriate
in the variational approach studied here. Although the precise values of $J_{ij}$ are to be determined,
that does not alter the qualitative physics we will discuss in this paper.
In the limit $U \rightarrow \infty$, the model is reduced to the t-J model.

We consider $\mid \Psi_L \rangle$ in Eq.(1) as a variational trial
wavefunction to examine the superconductor-
insulator transition at half filling, and to compare the gossamer superconducting state
with the RVB state away from the half filled.  In our theory,
$u_{\vec k}$, $v_{\vec k}$, and $\alpha$ are variational parameters.
In the limiting case $u_{\vec k} v_{\vec k} =0$, $\mid \Psi_{BCS}
\rangle$ reduces to the Fermi liquid state.
$\alpha=0$ corresponds to the uncorrelated state ($U=0$),
and $\alpha=1$ corresponds to the limit of no doubly occupied state.
Obviously $\alpha=1$ if $U \rightarrow \infty $.

The variational energy $E = \langle H \rangle$ is given by,
\begin{eqnarray}
E = Ud + \langle H_{t} \rangle + \langle H_J \rangle
\end{eqnarray}
where $d = \langle n_{i \uparrow} n_{i \downarrow} \rangle$
is the electron double occupation number, and $0 \leq d \leq 1$.
$\langle A \rangle$ is the expectation value (per site) of operator
$A$ in the state $\mid \Psi_L \rangle$.
The first term  is the intra-site Coulomb interaction energy,
while the second and the third terms are the average kinetic and
spin-spin correlation energies, respectively.  Note that in the Gutzwiller approach and at half filling,
$d$ is a measure of the mobile
carrier density $n_c$ and proportional to $n_c/m^*$ measured in the a.c. conductivity with
$m^*$ the effective mass.
At $d=0$, i.e. $\alpha=1$, we have $\langle H_{t} \rangle =0$, and  $E = \langle H_J \rangle$
at the half filling. This state describes a Mott insulator.
The case with $d > 0$ or $0 \leq \alpha < 1$ describes Laughlin's gossamer superconducting state.

The variational parameter $d$ is determined by the condition $\partial E(d)/\partial d =0$, or
\begin{eqnarray}
U + \partial \langle H_t \rangle / \partial d + \partial \langle H_J \rangle / \partial d = 0
\end{eqnarray}
At half filling, we expect a transition from the Mott insulator at larger $U$ to
the gossamer supercondutor at smaller $U$ as $U$ decreases passing through a critical point $U_c$.
To study this phase transition, we follow Brinkman and Rice ~\cite{brinkman} and compare the energies of the two states
with the difference that here
we consider the projected BCS state while Brinkman and Rice considered the projected Fermi liquid state.
The transition point $U_c$ is given by
$U_c = (-\partial \langle H_t \rangle /\partial d -  \partial \langle H_J \rangle /\partial d )_{\mid_{ d=0}}$.
For $U > U_c$, there is no solution of Eq. (4) for physical values of $d$, indicating  that $d=0$.

We use the Gutzwiller approximation~\cite{gutzwiller} to carry out the variation and to estimate $U_c$.
In the Gutzwiller approximation,
the effect of the partial projection operator on the doubly occupied sites is taken into account
by a classical statistical weighting factor which multiplies the quantum coherent result calculated for the unprojected state
$\mid  \Psi_{BCS}\rangle$. A clear description of the method has been given by Vollhardt~\cite{vollhardt}. The method was used
to study the 2-dimensional t-J model~\cite{zhang2}, where
the projection operator corresponds to the case
of $\alpha=1$ (the complete projection). In the present model, the
hopping and the spin-spin correlation energies in the state
$\mid \Psi_L \rangle$ are related to those
in the unprojected state $\mid \Psi_{BCS} \rangle$ by the corresponding renormalized constants $g_t$ and $g_s$:
\begin{eqnarray}
\langle c^{\dag}_{i\sigma}c_{j\sigma} \rangle = g_t \langle c^{\dag}_{i\sigma}c_{j\sigma} \rangle _{0} \nonumber \\
\langle \vec S_i \cdot  \vec S_j \rangle = g_s \langle \vec S_i \cdot  \vec S_j \rangle _0
\end{eqnarray}
where $\langle A \rangle_0$ is the expectation value of operator $A$ in the state $\mid \Psi_{BCS} \rangle$.
The renormalization facotrs $g_t$ and $g_s$ are determined by the ratios of the probabilities of the
corresponding physical processes in the
states $\mid\Psi_L \rangle$ and $\mid\Psi_{BCS} \rangle$. By counting the probabilities~\cite{zhang2}
we obtain these renormalization constants for the partially projected state ($n$: electron density),
\begin{eqnarray}
g_t = \frac{(n -2d)(\sqrt{ d} + \sqrt{1- n + d})^2}{(1- n/2)n} \nonumber \\
g_s = [\frac{(n - 2d)}{(1- n/2)n}]^2
\end{eqnarray}
The value of $g_t$ is the same as that  previously obtained for the projected Fermi liquid state ~\cite{gutzwiller,vollhardt}.
At $d=0$, we have $g_t =2\delta/(1+\delta)$ and $g_s = 4/(1+\delta)^2$, with $\delta =1 -n$,
recovering the results in Ref. \cite{zhang2}.
At half filling, $n=1$, we have $g_t = 8d(1-2d)$ and $g_s = 4(1-2d)^2$.  Using the Gutzwiller approximation,
The variational condition for $d$ in Eq. (4) becomes
\begin{eqnarray}
U + \frac{\partial g_t}{\partial d} \langle H_t \rangle_0 + \frac{\partial g_s}{\partial d} \langle H_J \rangle_0 = 0
\end{eqnarray}
The transition point between the Mott insulator and the gossamer superconductor is thus found to be,
$U_c = - 8 \langle H_t \rangle_0 + 16 \langle H_J \rangle_0$.
Note that $\langle H_t \rangle_0 < 0$, so that $U_c$ is generally positive if the kinetic energy term
in the uncorrelated state dominates. .
In the insulating phase, only the spin-spin interaction is non-zero. The problem becomes identical to that in the RVB theory
at half filling, and there is a redundancy in the fermion representation of the state due to a local SU(2) symmetry of the
spin Hamiltonian~\cite{affleck,zhang2}.
The redundancy is removed in the gossamer superconducting state for the kinetic energy term breaks the SU(2) symmetry,
similar to the effect of doping in the t-J model. At $d <<1$, the symmetry of the gossamer superconductivity
is the same as the symmetry of the RVB state.
Within the Gutzwiller approximation, the pairing order parameter in the gossamer superconductor is related to the
uncorrelated state by a renormalized factor
$g_t$,
\begin{eqnarray}
\langle c_{\vec k \uparrow}c_{-\vec k \downarrow} \rangle = g_t \langle c_{\vec k \uparrow}c_{-\vec k \downarrow} \rangle_0
\end{eqnarray}
Near the transition point, $g_t = 8d << 1$, indicating the smallness of the superfluid density, a quantitative
measure of the gossamer superconductivity. It is interesting to note that the pairing order parameter in the RVB
state has also the form of Eq. (8) with $g_t = 2\delta$ for $\delta <<1$. This comparison indicates that
a gossamer superconductor with double occupation $d$
at half filling is similar to the RVB superconductor at doping $\delta$ with the correspondance of $\delta = 4d$.

In what follows we take an example and consider the effective Hamiltonian in a 2-dimensional square lattice
with only the nearest neighbour hopping $t_{ij} = t$
and the nearest neighbor spin coupling
$J_{ij} =J$ and consider  the case $n \leq 1$.
For any given value of $d$,  the Coulomb interaction term in the present theory contributes
 a constant $Ud$ to the variational energy, and
the variational procedure for other parameters ($u_{\vec k}$ and $v_{\vec k}$)
is almost the same as that in study of the t-J model
carried out in Ref. {\cite{zhang2}
except that
the renormalization constants $g_t$ and $g_s$ here depend also on the double occupation $d$.

We introduce two correlation functions ($\tau = x, y$),
$\Delta_{\tau} = \sum_{\sigma}\sigma\langle c_{i \sigma}c_{i + \tau, -\sigma}\rangle _0$,
$\chi_{\tau} = \sum_{\sigma}\langle c^{\dag}_{i \sigma}c_{i + \tau, \sigma} \rangle_0$.
The variational solution is then given by the coupled gap equations,
\begin{eqnarray}
\Delta_{\tau} = \sum_{\vec k} \cos{k_{\tau}} \, \Delta_{\vec k}/E_{\vec k} \nonumber \\
\chi_{\tau} = - \sum_{\vec k} \cos{k_{\tau}} \, \chi_{\vec k}/E_{\vec k}  \nonumber
\end{eqnarray}
where
$\Delta_{\vec k} = \sum_{\tau}\Delta_{\tau} \cos{k_{\tau}}$,
$\chi_{\vec k} = \tilde{\epsilon_{\vec k}} - \sum_{\tau}\chi_{\tau} \cos{k_{\tau}}$.
In the above equations, $E_{\vec k} = \sqrt{\mid \Delta_{\vec k}\mid ^2 + \chi_{\vec k}^2}$,
$\tilde{\epsilon_{\vec k}} = [- 2g_t t (\cos{k_x} + \cos{k_y})- \tilde{\mu}]/(3g_s J/4)$,
and $\tilde{\mu}$ is related to the chemical potential $\mu$ by
\begin{eqnarray}
\mu = \tilde{\mu} + \frac{\partial g_t}{\partial n} \langle H_t \rangle_0 +
\frac{\partial g_s}{\partial n}\langle H_J \rangle_0
\end{eqnarray}
In Eq. (9), the second and the third terms originate from the $n$ dependences of $g_t$ and $g_s$
in the variational procedure~\cite{zhang2}, which will be important in calculation of the chemical potential
of the state.
These gap equations must be solved simultaneously with Eq. (7) for  $d$ and the electron number equation
given by
$\delta = \sum_{\vec k} \chi_{\vec k}/E_{\vec k}$.

We first discuss the half filled case.
The ground state of the insulating phase ($d=0$) is the same as that of the Heisenberg model.
In the metallic phase ($0 < d << 1$), the kinetic energy breaks the local SU(2)
symmetry and favors the d-wave superconducting state with $\Delta_x = - \Delta_y$.
The symmetry is the same as the symmetry studied in the
t-J model slighlt away from the half filled\cite{kotliar,fukuyama,zhang2}.
At the superconductor- insulator transition point,
we have $\langle H_t \rangle_0 = -2\sqrt{2}t C$, and
$\langle H_J \rangle _0 = -(3/4)JC^2$, with
$C = \frac{1}{2}\sum_{\vec k}\sqrt{\cos^2{k_x} + \cos^2{k_y}} = 0.479$.
We estimate from these values that $U_c = 10.8 t - 2.75 J $.

We now discuss the slightly less than half filled case. We expect that
the variational parameter $d$ is a smooth function of the electron density around the half filled.
The gossamer superconducting state essentially remains unchanged in the regime $\delta << d$,
and the superconducting order parameter is mainly controlled by $d$, weakly depending on $\delta$ as we can see
from the expression for $g_t$.  The chemical potential $\mu$ can be calculated by using Eq. (9).
In the limit $\delta \rightarrow 0^+$, $\tilde{\mu} \rightarrow 0$, and we have
$\mu \rightarrow  - 4(1- 4d) \langle H_t \rangle_0 + 8(1-2d) \langle H_J \rangle_0 =U/2$. In the last step of
the above calculations,
we have used varirational equation (7) to relate the kinetic and spin coupling energies to the Coulomb energy.
Since $\mu =U/2$ at the half filled by electron-hole symmetry
of the model, we conclude that in the gossamer superconducting state
the chemical potential is continuous at the half filled, and is pinned at the mid of the
lower and the higher Hubbard bands.  This result is reasonable because the gossamer superconducting state
is a metallic state and the chemical potential is expected to be continuous~\cite{laughlin}.
This feature is in contrast to the RVB state
discussed  below.

If  $U > U_c$ with $U_c$ defined as the critical $U$ at half filling, the state
changes dramatically
from an insulator to a RVB superconducting state as the electron density varies away from the half filled.
At $U >> U_c$, $d$ changes very little from zero~\cite{d-value}, the physics is essentially the same as
that given by the t-J model. While the RVB state is similar to the gossamer superconducting
state in the sense that they have the same pairing symmetries and  small pairing order parameters,
the chemical potential in the RVB state is very different from that in the gossamer superconductor.
To see this explicitly, we consider the limit $\delta \rightarrow 0^+$, so that we have
$\tilde{\mu} \rightarrow 0$, and
$\partial g_t/\partial n = - 2$,
$\partial g_J /\partial n =  8$.  From these values, we obtain
$\mu \rightarrow -2 \langle H_t \rangle_0 + 8 \langle H_J \rangle_0 \approx 2.7 t - 1.38 J$, which is
about $U_c/4$ for $J << t$, and is much smaller than the chemical potential $U/2$ at the exact half filled.
We conclude that the chemical potential in the RVB state is discontinuous at the half filled,
and it is shifted from $U/2$ at the half filled to the lower Hubbard band away from the half filled.
The difference in the chemical potentials in the gossamer and the RVB states
can in principle be distinguished
in spectroscopic experiments, although other symmetry broken states not included in the variational
theory and the inhomogeneity will complicate the analyses..

The actual ground state of the Hubbard model in a square lattice and at half filled
is  an antiferromagnet, which we have not included
in our variational wavefunction. Nevertheless, the phase transition and the similarities and the
differences between the gossamer and the RVB superconductors should be relevant
to the systems away from half filled.  It is interesting to note that
the half filled superconducting state may be stabilized against
the antiferromagnetism in the presence of an explicit pair hopping term in the Hubbard model
as studied by Assaad et al. \cite{assaad}.
The gossamer superconductor should also be relevant to
the frustrated magnetic systems where the anitferromagnetism
is suppressed.  For instance, there have been numerical studies of
the antiferromagnetic Heisenberg spin model with nearest and next nearest neighbor
couplings to show evidences for
a spin liquid ground state~\cite{sorella}.  There have been numerical studies of
the Hubbard model at half filled with the nearest and next nearest
hoppings to demonstrate a paramagnetic insulating phase with a transition to a metallic phase at zero temperature
as $U$ decreases\cite{kondo,imada}. It will be interesting to examine
the possible gossamer superconductivity in that metallic phase.

In summary, we have used Gutzwiller variational method to study an effective Hubbard model.
The calculation based on the Gutzwiller approximation
supports Laughlin's recent proposal of gossamer superconductor at relatively smaller
intra-site electron Coulomb repulsion $U$, and predicts a phase transition from the gossamer superconductor to the Mott
insulator as $U$ increases at density 1 electron per site.
The gossamer superconductor is similar to the RVB superconducting state
with the major difference on the positions of their chemical potentials.
The Gutzwiller approximation we used in this paper has been previously tested against variational Monte Carlo
method ~\cite{gros} with quite good agreement~\cite{zhang2}. We believe that the qualitative
conclusions obtained here should be reliable, while the quantitative values should be taken with care and may be refined
using variational Monte Carlo or other numerical calculations. .

We would like to acknowledge many useful and stimulate discussions with R. Laughlin, T. M. Rice, C. Honerkamp,
and Y. Yamashita. This work was partially supported by the US DOE grant DE/FG03-01ER45687. Work at LANL was
supported by the U.S. Department of Energy.


\begin{references}
\bibitem{d-value}  The value of $d$ in the doped Mott insulating phase can be estimated in our theory, and is
very small in the limit $U >> U_c$. For small $\delta$ and $J << t$, we find $d \approx \delta (U_c/4U)^2 << \delta$
from equations (5) -(7).
\end{references}
\end{document}